\pdfoutput=1
\documentclass[english,fleqn,twoside,espcrc2,floatfix,letterpaper]{article}
\usepackage[T1]{fontenc}
\usepackage[latin1]{inputenc}
\usepackage{graphicx}
\usepackage{amssymb}
\usepackage[numbers]{natbib}

\providecommand{\tabularnewline}{\\}

\usepackage[headings]{espcrc2}

\readRCS
$Id: espcrc2.tex,v 1.2 2004/02/24 11:22:11 spepping Exp $
\title{Nuclear Parton Distribution Functions\thanks{To 
appear in the proceedings of  the 
Ringberg Workshop, 
New Trends in HERA Physics 2008; 
October 5 --- 10, 2008, 
Ringberg Castle, Tegernsee.
}}

\author{
I.~Schienbein,\address{LPSC, Universit\'e Joseph Fourier Grenoble 1,\\
CNRS/IN2P3, Institut National Polytechnique de Grenoble, 38026 Grenoble, France}
J.~Y.~Yu,\address[SMU]{Southern Methodist University, Dallas, TX 75206, USA}
C.~Keppel,\address[jlab]{Thomas Jefferson National Accelerator Facility, Newport News, VA 23602, USA}\address{Hampton University, Hampton, VA, 23668, USA} 
J.~G.~Morfin,\address{Fermilab, Batavia, IL 60510, USA}
F.~Olness,${}^b$\thanks{Presented by Fred Olness.}
and
J.F.~Owens,\address[FSU]{Florida State University, Tallahassee, FL 32306-4350, USA}
}

\runtitle{Nuclear PDFs}
\runauthor{Schienbein,  {\it et al.}}

\def\lsim{\mathrel{\hbox{\rlap{\hbox{\lower4pt\hbox{$\sim$}}}\hbox{$<$}}}}
\def\gsim{\mathrel{\rlap{\lower4pt\hbox{\hskip1pt$\sim$}} \raise1pt\hbox{$>$}}}
\newcommand{\gev}{\ensuremath{\mathrm{GeV}}}
\newcommand{\gevsq}{\ensuremath{\mathrm{GeV^2}}}

\usepackage{babel}

\begin{document}
\begin{abstract}
We study nuclear effects of charged current deep inelastic neutrino-iron scattering in the framework of a $\chi^2$ analysis of parton distribution functions (PDFs). We extract a set of iron PDFs which are used to compute $x_{Bj}$-dependent and $Q^2$-dependent nuclear correction factors for iron structure functions which are required in global analyses of free nucleon PDFs.  We compare our results with nuclear correction factors from neutrino-nucleus scattering models and correction factors for $\ell^\pm$-iron scattering.  We find that, except for very high $x_{Bj}$, our correction factors differ in both shape and magnitude from the correction factors of the models and charged-lepton scattering.

\end{abstract}
\maketitle

\def\gevsq{{\rm GeV}^2}
\def\gev{{\rm GeV}}
\def\eqref#1{\ref{#1}}
\def\overset#1#2{{#1 \atop #2}}

\section{Impact of Nuclear Corrections on PDFs }

The high statistics measurements of neutrino deeply inelastic scattering
(DIS) on heavy nuclear targets has generated significant interest
in the literature since these measurements provide valuable information
for global fits of parton distribution functions (PDFs). It is necessary
to use both Charged Current (CC) $W^{\pm}$ probes and Neutral Current
(NC) $\{\gamma,Z\}$ probes to disentangle the separate PDF flavor
components. Toward this goal, the use of nuclear targets is unavoidable
due to the weak nature of the $\{W^{\pm},Z\}$ interactions, and this
complicates the extraction of free nucleon PDFs because model-dependent
corrections must be applied to the data. 

In early PDF analyses, the nuclear corrections were static correction
factors without any (significant) dependence on the energy scale $Q$,
the atomic number $A$, or the specific observable. The increasing
precision of both the experimental data and the extracted PDFs demand
that the applied nuclear correction factors be equally precise as
these contributions play a crucial role in determining the PDFs. 

In this study we reexamine the source and size of the nuclear corrections
that enter the PDF global analysis, and quantify the associated uncertainty.
Additionally, we provide the foundation for including the nuclear
correction factors as a dynamic component of the global analysis so
that the full correlations between the heavy and light target data
can be exploited.

A recent study\citep{Owens:2007kp} analyzed the impact of new data
sets from the NuTeV, Chorus, and E-866 Collaborations on the PDFs.
This study found that the NuTeV data set (together with the model
used for the nuclear corrections) pulled against several of the other
data sets, notably the E-866, BCDMS and NMC sets. Reducing the nuclear
corrections at large values of $x$ reduced the severity of this pull
and resulted in improved $\chi^{2}$ values. These results suggest
on a purely phenomenological level that the appropriate nuclear corrections
for $\nu$-DIS may well be smaller than assumed.

\section{Global Analysis Framework}

To investigate this question further, we use the high-statistics $\nu$-DIS
experiments to perform a dedicated PDF fit to neutrino--iron data.\citep{Schienbein:2007fs}
Since we first will study iron alone and will not (initially) combine
the data with measurements on different target materials, we need
not make any assumptions about the nuclear corrections; this side-steps
a number of difficulties.\citep{Gomez:1993ri,Owens:2007kp,Thorne:2006zu}
While this approach has the advantage that we do not need to model
the $A$-dependence, it has the drawback that the data from just one
experiment will not be sufficient to constrain all the parton distributions;
therefore, other assumptions must enter the analysis. The theoretical
framework will roughly follow the CTEQ6 analysis of free proton PDFs.\citep{Pumplin:2002vw}
We outline the key features of our analysis below, and focus on the
issues specific to our study of NuTeV neutrino--iron data in terms
of nuclear parton distribution functions.

\subsection{Basic formalism}

For our PDF analysis, we will use the general features of the QCD-improved
parton model and the $\chi^{2}$ analyses as outlined in Ref.~\citep{Pumplin:2002vw}.
We adopt the framework of the recent CTEQ6 analysis of proton PDFs
where the input distributions at the scale $Q_{0}=1.3~\gev$ are parameterized
as 

\[
xf_{i}(x,Q_{0})=A_{0}x^{A_{1}}(1-x)^{A_{2}}e^{A_{3}x}(1+e^{A_{4}}x)^{A_{5}}\]
for $i=\{u_{v},d_{v},g,\bar{u}+\bar{d},s,\bar{s}\}$, and 

\[
xf_{i}(x,Q_{0})=A_{0}x^{A_{1}}(1-x)^{A_{2}}+(1+A_{3}x)(1-x)^{A_{4}}\]
for $i=\{\bar{d}/\bar{u}\}$ where $u_{v}$ and $d_{v}$ are the up-
and down-quark valence distributions, $\bar{u}$, $\bar{d}$, $s$,
$\bar{s}$ are the up, down, strange and anti-strange sea distributions,
and $g$ is the gluon. Furthermore, the $f_{i}=f_{i}^{p/A}$ denote
parton distributions of \emph{bound protons in the nucleus $A$},
and the variable $0\le x\le A$ is defined as $x:=Ax_{A}$ where $x_{A}=Q^{2}/2p_{A}\cdot q$
is the usual Bjorken variable formed out of the four-momenta of the
nucleus and the exchanged boson. This parameterization is designed
for $0\le x\le1$ and we here neglect%
\footnote{While the nuclear PDFs can be finite for $x>1$, the magnitude of
the PDFs in this region is negligible for the purposes of the present
study (\textit{cf.}, Refs.~\protect\citep{Hirai:2001np,Hirai:2004wq,Hirai:2007sx,Eskola:1998df,Eskola:2007my}). %
} the distributions at $x>1$. Note that the condition $f_{i}(x>1,Q)=0$
is preserved by the DGLAP evolution and has the effect that the evolution
equations and sum rules for the $f_{i}^{p/A}$ are the same as in
the free proton case.

The PDFs for a nucleus $(A,Z)$ are constructed as \[
f_{i}^{A}(x,Q)=\frac{Z}{A}\ f_{i}^{p/A}(x,Q)+\frac{(A-Z)}{A}\ f_{i}^{n/A}(x,Q)\]
 where we relate the distributions inside a bound neutron, $f_{i}^{n/A}(x,Q)$,
to the ones in a proton by assuming isospin symmetry. The nuclear
structure functions are given by parallel relations such that they
can be computed in next-to-leading order as convolutions of the nuclear
PDFs with the conventional Wilson coefficients, \textit{i.e.}, generically
$F_{i}^{A}(x,Q)=\sum_{k}C_{ik}\otimes f_{k}^{A}$.

In order to take into account heavy quark mass effects we calculate
the relevant structure functions in the ACOT scheme\citep{Aivazis:1993kh,Aivazis:1993pi}
in NLO QCD.\citep{Kretzer:1998ju} Finally, the differential cross
section for charged current (anti-)neutrino--nucleus scattering is
given in terms of three structure functions: \begin{eqnarray*}
\frac{d^{2}\sigma}{dx\, dy}^{\overset{(-)}{\nu}A} & = & \frac{G_{F}^{2}ME}{\pi}\left[(1-y-\frac{Mxy}{2E})F_{2}^{\overset{(-)}{\nu}A}\right.\\
 & + & \left.\frac{y^{2}}{2}2xF_{1}^{\overset{(-)}{\nu}A}\pm y(1-\frac{y}{2})xF_{3}^{\overset{(-)}{\nu}A}\right]\,,\end{eqnarray*}
 where the '$+$' ('$-$') sign refers to neutrino (anti-neutrino)
scattering and where $G_{F}$ is the Fermi constant, $M$ the nucleon
mass, and $E$ the energy of the incoming lepton (in the laboratory
frame).

\subsection{Methodology \label{method}}

The basic formalism described in the previous sections is implemented
in a global PDF fitting package, \textit{but} with the difference
that no nuclear corrections are applied to the analyzed data; hence,
the resulting PDFs are for a bound proton in an iron nucleus. The
parameterization provides enough flexibility to describe current data
sets entering a global analysis of free nucleon PDFs; given that the
nuclear modifications of the $x$-shape appearing in this analysis
are modest, this parameterization will also accommodate the iron PDFs.

Because the neutrino data alone do not have the power to constrain
all of the PDF components, we will need to impose some minimal set
of external constraints. For example, our results are rather insensitive
to the details of the gluon distribution with respect to both the
overall $\chi^{2}$ and also the effect on the quark distributions.
The nuclear gluon distribution is very weakly constrained by present
data, and a gluon PDF with small nuclear modifications has been found
in the NLO analysis of Ref.~\citep{deFlorian:2003qf}. We have therefore
fixed the gluon input parameters to their free nucleon values. For
the same reasons the gluon is not sensitive to this analysis, fixing
the gluon will have minimal effect on our results. Furthermore, we
have set the $\bar{d}/\bar{u}$ ratio to the free nucleon result assuming
that the nuclear modifications to the down and up sea are similar
such that they cancel in the ratio. This assumption is supported by
Fig.~6 in Ref.~\citep{deFlorian:2003qf}.

\section{Analysis of iron data \label{sec:iron}}

\begin{figure}[!t]
\includegraphics[width=0.95\columnwidth]{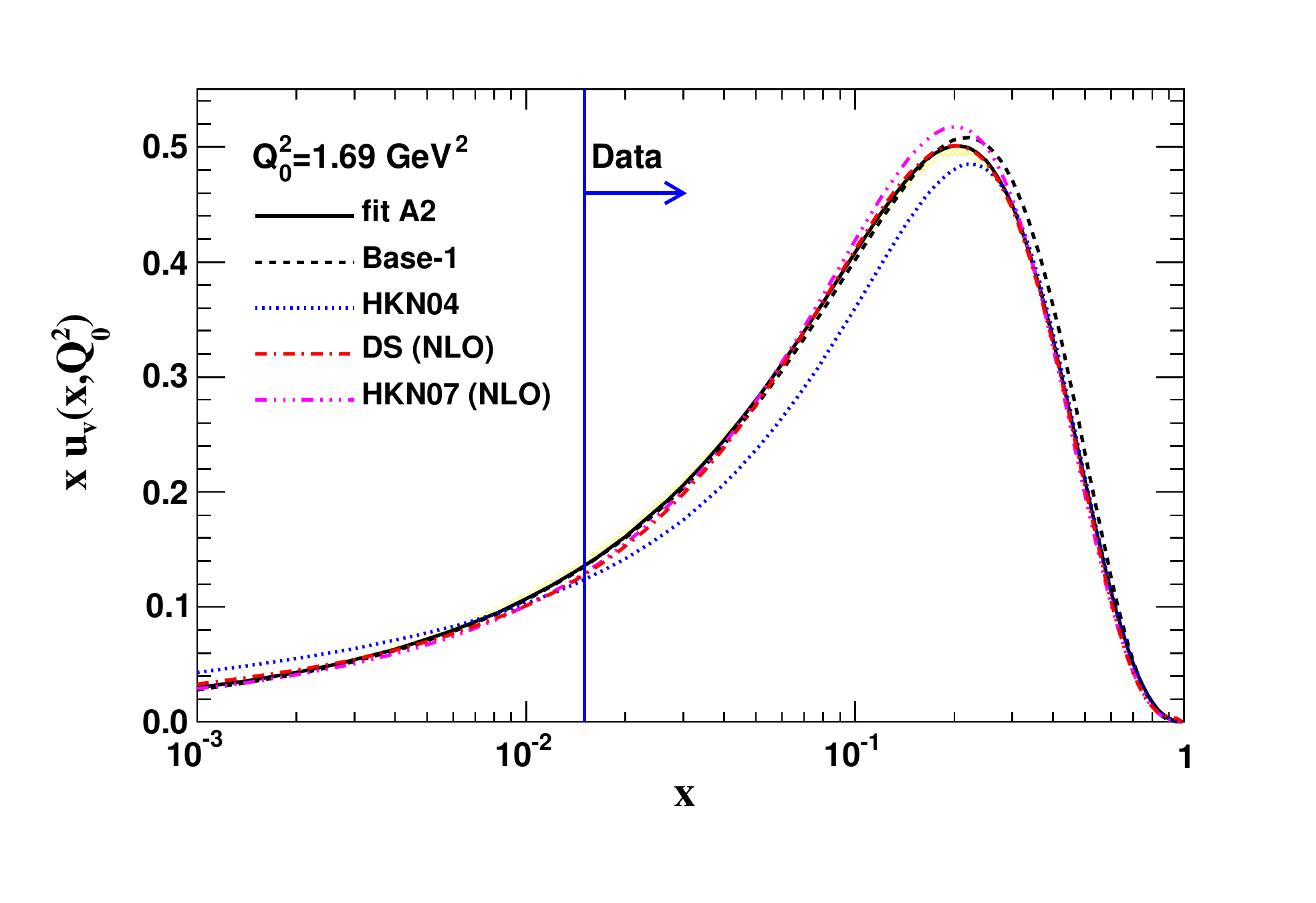}\vspace{-1cm}

\includegraphics[width=0.95\columnwidth]{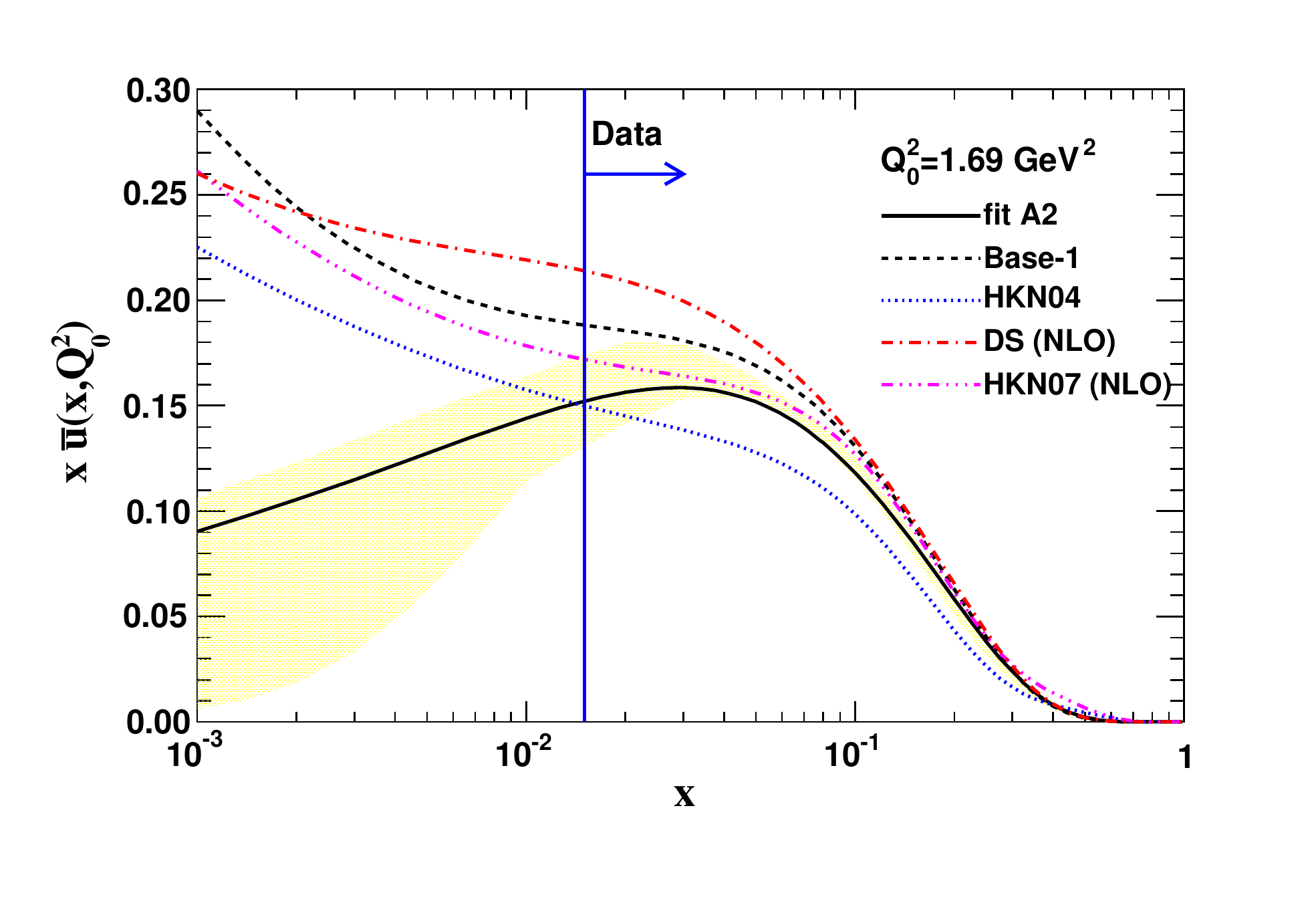}\vspace{-1cm}

\includegraphics[width=0.95\columnwidth]{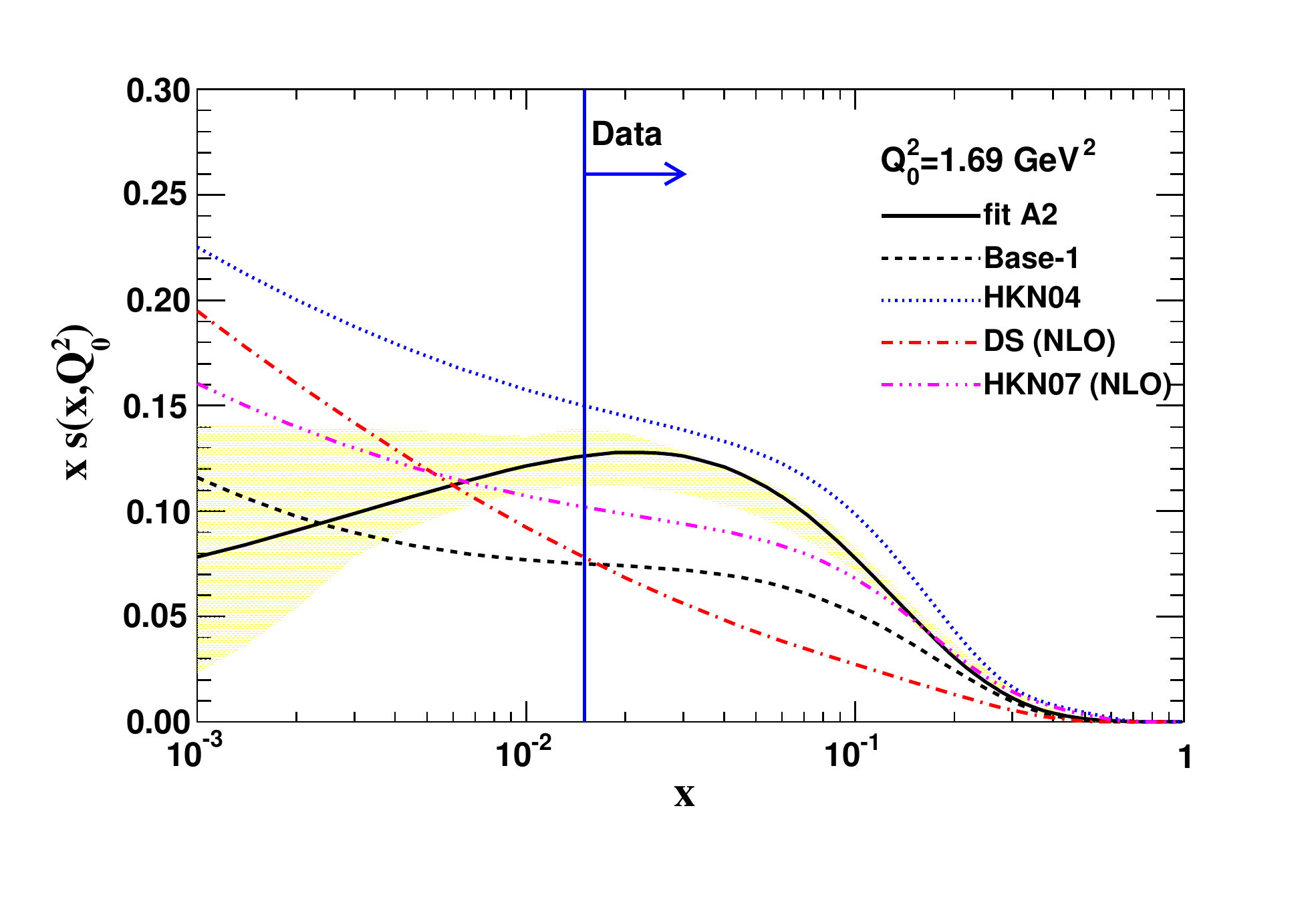}\vspace{-1cm}

\caption{Parton distributions for iron. The central PDF from fit `A2' is shown
by the solid line. The dashed lines depict parton distributions constructed
with $A=56$ and $Z=26$ using the \hbox{Base-1} free-proton PDFs.
Additional results are shown from HKN04 \citep{Hirai:2004wq}, (NLO)
HKN07 \citep{Hirai:2007sx}, and (DS) \citep{deFlorian:2003qf}. The
vertical line marks the lower limit of the data in the $x$ variable.\label{fig:pdf}}
 
\end{figure}

\subsection{Iron Data Sets}

We determine iron PDFs using the recent NuTeV differential neutrino
and anti-neutrino DIS cross section data.\citep{Tzanov:2005kr} In
addition, we include NuTeV/CCFR dimuon data \citep{Goncharov:2001qe}
which are sensitive to the strange quark content of the nucleon. There
are other measurements of neutrino--iron DIS available in the literature
from the CCFR \citep{Oltman:1992pq,Seligman:1997mc,Yang:2000ju,Yang:2001rm},
CDHS\citep{Abramowicz:1984yk} and CDHSW\citep{Berge:1989hr} collaborations;
see, e.g., Ref.\ \citep{Conrad:1997ne} for a review. There is also
a wealth of charged lepton--iron DIS data including SLAC\citep{Dasu:1993vk}
and EMC\citep{Aubert:1986yn,Aubert:1987da}.%
\footnote{\textit{Cf.} the Durham HEP Databases for a complete listing: \texttt{http://www-spires.dur.ac.uk/hepdata/} %
} For the initial study we limit our analysis to the NuTeV experiment
alone; we will compare and contrast different experiments in a subsequent
study.

\subsubsection{PDF Reference Sets}

For the purposes of this study, we use two different reference sets
of free-proton PDFs which we denote \hbox{`Base-1'} and \hbox{`Base-2'.}

Since we focus on iron PDFs and the associated nuclear corrections,
we need a base set of PDFs which are essentially free of any nuclear
effects; this is the purpose of the Base-1 reference set \citep{Owens:2007kp}.
Therefore, to extract the Base-1 PDFs we omit the CCFR and NuTeV data
from our fit so that our base PDFs do not contain any large residual
nuclear corrections.%
\footnote{We do retain the deuteron data as this has only a small correction
over the central $x$-range,\citep{Gomez:1993ri,Owens:2007kp}. The
deuteron correction has been applied in the Base-1 fit. Also, for
the Drell-Yan Cu data (E605), the expected nuclear corrections in
this kinematic range are small (a few percent) compared to the overall
normalization uncertainty (15\%) and systematic error (10\%). %
} The absence of such nuclear effects will be important when we extract
the nuclear corrections factors.

The Base-2 PDFs are essentially the CTEQ6.1M PDFs with a modified
strange PDF introduced to accommodate the NuTeV dimuon data.%
\footnote{These PDFs have been determined from a fit to the same data set as
in the CTEQ6 analysis with the addition of the NuTeV dimuon data.
The changes to the strange sea induce only minor changes to the other
fit parameters; this has a minimal effect on the particular observables
($d\sigma$, $F_{2}$) we examine in the present study. %
} In the manner of the CTEQ6.1M PDF's, the Base-2 fit does not apply
any deuteron corrections to the data; this is in contrast to the Base-1
PDFs. Also, the Base-2 fit does include the CCFR data that has been
corrected to a free nucleon using charged-lepton correction factors.\citep{Yang:2001rm}

By comparing the free-proton PDF \hbox{`Base-1'} and \hbox{`Base-2'}
sets with the iron PDF sets, we can gauge the size of the nuclear
effects. Furthermore, differences between observables using the \hbox{`Base-1'}
respectively the \hbox{`Base-2'} reference sets will indicate the
uncertainty due to the choice of the free-proton PDF.

\subsubsection{Comparison of the Fits with Data}

Specifically, we determine iron PDFs using the recent NuTeV differential
neutrino (1371 data points) and anti-neutrino (1146 data points) DIS
cross section data,\citep{Tzanov:2005kr} and we include NuTeV/CCFR
dimuon data (174 points) which are sensitive to the strange quark
content of the nucleon. Using the ACOT scheme, we impose kinematic
cuts of $Q>2$~GeV and $W>3.5$~GeV, and obtain a good fit with
a $\chi^{2}$ of 1.35 per data point; we identify this fit as `A2.'\citep{Schienbein:2007fs}

\subsection{Iron PDFs}

We now examine the nuclear (iron) parton distributions $f_{i}^{A}(x,Q^{2})$
in Figure~\ref{fig:pdf} which shows the PDFs from fit `A2' at our
input scale $Q_{0}=m_{c}=1.3~\gev$ versus $x$. For an almost isoscalar
nucleus like iron the $u$ and $d$ distributions are very similar.
Therefore, we only show the $u_{v}$ and $\bar{u}$ partons, together
with the strange sea.%
\footnote{While iron is roughly isoscalar, other nuclear PDFs can exhibit larger
differences between the $u$ and $d$ distributions---the extreme
case being the free-proton PDF. When comparing PDFs we must keep in
mind that it is ultimately the structure functions which are the physical
observables. %
} As explained above, the gluon distribution is very similar to the
familiar CTEQ6M gluon at the input scale such that we don't show it
here. In order to indicate the constraining power of the NuTeV data,
the band of reasonable fits is depicted. The fits in this band were
obtained (as outlined above) by varying the initial conditions and
the number of free parameters to fully explore the solution space.
All the fits shown in the band have $\chi^{2}/DOF$ within 0.02, which
roughly corresponds to a range of $\Delta\chi^{2}\sim50$ for the
2691 data points.

As can be seen in Figure~\ref{fig:pdf}, the $u_{v}$ distribution
has a very narrow band across the entire $x$-range. The up- and strange-sea
distributions are less precisely determined. At values of $x$ down
to, say, $x\simeq0.07$ the bands are still reasonably well confined;
however, they open up widely in the small-$x$ region. Cases where
the strange quark sea lies above the up-quark sea are unrealistic,
but are present in some of the fits since this region ($x\lesssim0.02$)
is not constrained by data. We have included the curves for our free-proton
\hbox{Base-1} PDFs (dashed), as well as the HKN04 \citep{Hirai:2004wq}
(dotted), the NLO HKN07 \citep{Hirai:2007sx} (dotted-dashed), and
DS \citep{deFlorian:2003qf} (dot-dashed) nuclear PDFs.%
\footnote{In a recent publication, Eskola \textit{et al.} \citep{Eskola:2007my}
perform a global reanalysis of their ESK98 \citep{Eskola:1998df}
nuclear PDFs. While we do not present a comparison here, the results
are compatible with those distributions displayed in Fig.~\ref{fig:pdf};
a comparison can be found in Figs.~10 and 11 of Ref.~\citep{Eskola:2007my}. %
}

The comparison with the Base-1 PDFs is straightforward since the same
theoretical framework (input scale, functional form, NLO evolution)
has been utilized for their determination. Therefore, the differences
between the solid band (`A2') and the dashed line (\hbox{Base-1})
exhibit the nuclear effects, keeping in mind that the free-proton
PDFs themselves have uncertainties.

For the comparison with the HKN04 distributions, it should be noted
that a SU(3)-flavor symmetric sea has been used; therefore, the HKN04
strange quark distribution is larger, and the light quark sea smaller,
than their Base-1 PDF counterparts over a wide range in $x$. Furthermore,
the HKN04 PDFs are evolved at leading order.

In a recent analysis, the HKN group has published a new set of NPDFs
(HKN07) including uncertainties \citep{Hirai:2007sx}. They provide
both LO and NLO sets of PDFs, and we display the NLO set. These PDFs
also use a more general set of sea distributions such that $\bar{u}(x)\not=\bar{d}(x)\not=\bar{s}(x)$
in general.

The DS PDFs are linked to the GRV98 PDFs \citep{Gluck:1998xa} with
a rather small radiatively generated strange sea distribution. Consequently,
the light quark sea is enhanced compared to the other sets. Additionally,
the DS sets are evolved in a 3-fixed-flavor scheme in which no charm
parton is included in the evolution. However, at the scale $Q=m_{c}$
of Fig.~\ref{fig:pdf} this is of no importance.

\section{Nuclear Correction Factors \label{sec:nfac}}

In the previous section we analyzed charged current $\nu$--Fe data
with the goal of extracting the iron nuclear parton distribution functions.
In this section, we now compare our iron PDFs with the free-proton
PDFs (appropriately scaled) to infer the proper heavy target correction
which should be applied to relate these quantities.

Within the parton model, a nuclear correction factor $R[{\cal O}]$
for an observable ${\cal O}$ can be defined as follows: \begin{equation}
R[{\cal O}]=\frac{{\cal O}[{\rm NPDF}]}{{\cal O}[{\rm free}]}\label{eq:R}\end{equation}
 where ${\cal O}[{\rm NPDF}]$ represents the observable computed
with nuclear PDFs, and ${\cal O}[{\rm free}]$ is the same observable
constructed out of the free nucleon PDFs. Clearly, $R$ can depend
on the observable under consideration simply because different observables
may be sensitive to different combinations of PDFs.

This means that the nuclear correction factor $R$ for $F_{2}^{A}$
and $F_{3}^{A}$ will, in general, be different. Additionally, the
nuclear correction factor for $F_{2}^{A}$ will yield different results
for the charged current $\nu$--$Fe$ process ($W^{\pm}$ exchange)
as compared with the neutral current $\ell^{\pm}$--$Fe$ process
($\gamma$ exchange). Schematically, we can write the nuclear correction
for the DIS structure function $F_{2}$ in a charged current (CC)
$\nu$--$A$ process as:%
\footnote{The corresponding anti-neutrino process is obtained with a $u\leftrightarrow d$
interchange.%
} 

\global\long\def\p{{\,\emptyset}}
 \begin{eqnarray}
R_{CC}^{\nu}(F_{2};x,Q^{2}) & \simeq & \frac{d^{A}+\bar{u}^{A}+...}{d^{\p}+\bar{u}^{\p}+...}\label{eq:rcc}\end{eqnarray}
 and contrast this with the neutral current (NC) $\ell^{\pm}$--$A$
process:\begin{eqnarray*}
R_{NC}^{e,\mu}(F_{2};x,Q^{2})\simeq\qquad\qquad\qquad\qquad\qquad\qquad\\[5pt]
\frac{\left(-\frac{1}{3}\right)^{2}\left[d^{A}+\bar{d}^{A}+...\right]+\left(+\frac{2}{3}\right)^{2}\left[u^{A}+\bar{u}^{A}+...\right]}{\left(-\frac{1}{3}\right)^{2}\left[d^{\p}+\bar{d}^{\p}+...\right]+\left(+\frac{2}{3}\right)^{2}\left[u^{\p}+\bar{u}^{\p}+...\right]}\ ,\end{eqnarray*}
where the superscript {}``$\p$'' denotes the {}``free nucleon''
PDF which is constructed via the relation: \begin{equation}
f_{i}^{\p}(x,Q)=\frac{Z}{A}\ f_{i}^{p}(x,Q)+\frac{(A-Z)}{A}\ f_{i}^{n}(x,Q)\quad.\label{eq:pdf2}\end{equation}
 Clearly, the $R$-factors depend on both the kinematic variables
and the factorization scale. Finally, we note that Eq.~(\eqref{eq:R})
is subject to uncertainties of both the numerator and the denominator.

We will now evaluate the nuclear correction factors for our extracted
PDFs, and compare these with selected results from the literature
\citep{Kulagin:2004ie,Kulagin:2007ju}.%
\footnote{Note that our comparison with the Kulagin--Petti model is based on
the work in Ref.~\citep{Kulagin:2004ie}.%
} Because we have extracted the iron PDFs from only iron data, we do
not assume any particular form for the nuclear $A$-dependence; hence
the extracted $R[{\cal O}]$ ratio is essentially model independent.

\subsection{$F_{2}^{Fe}/F_{2}^{D}$ NC charged lepton scattering}

\begin{figure}[t]
\includegraphics[width=0.95\columnwidth]{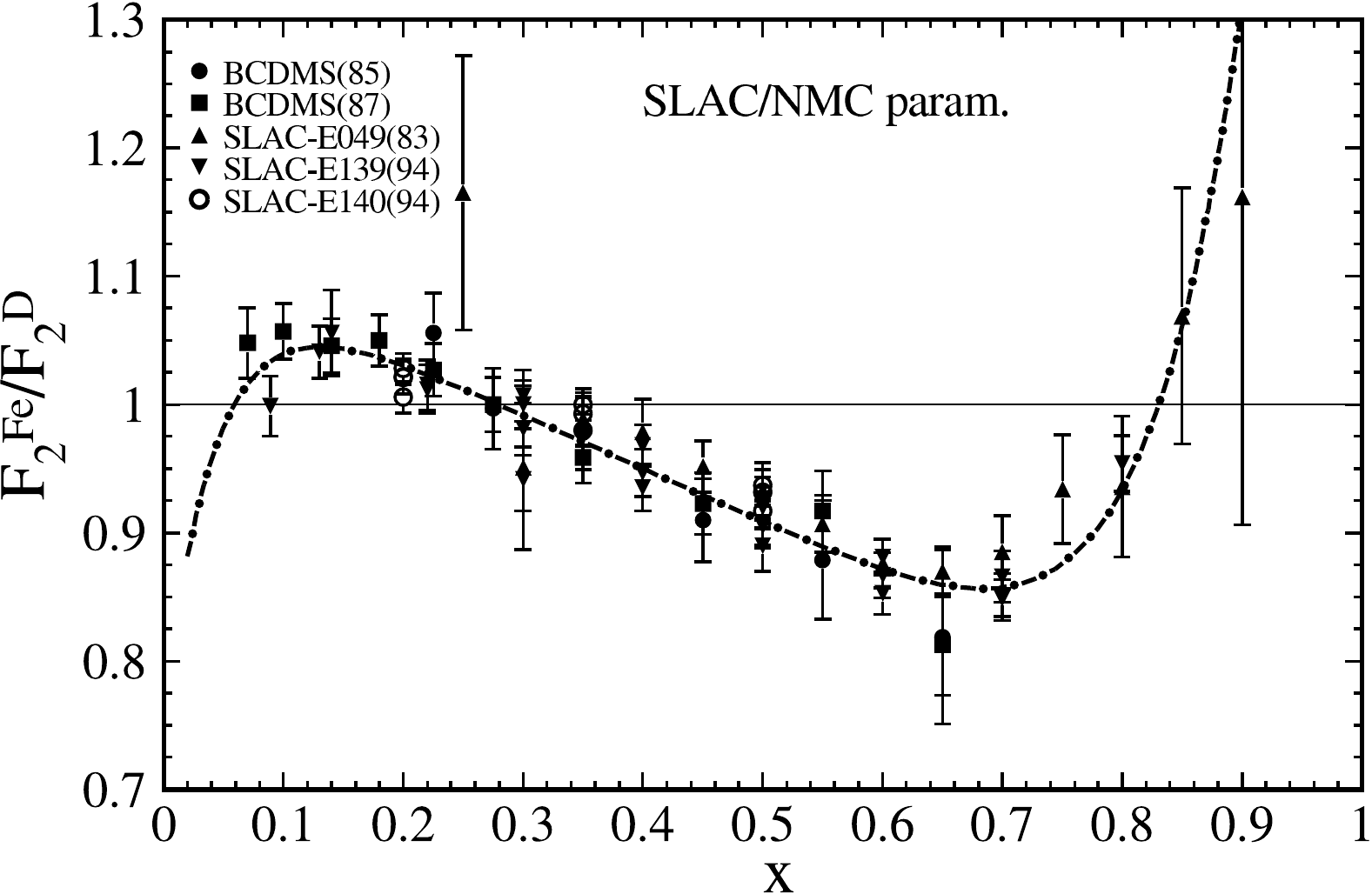}

\vspace{-0.5cm}
\caption{Parameterization for the neutral current charged lepton structure
function $F_{2}^{Fe}/F_{2}^{D}$. For comparison we show experimental
results from the BCDMS collaboration (BCDMS-85 \citep{Bari:1985ga},
BCDMS-87 \citep{Benvenuti:1987az}) and from experiments at SLAC (SLAC-E049
\citep{Bodek:1983qn}, SLAC-E139 \citep{Gomez:1993ri}, and SLAC-E140
\citep{Dasu:1993vk}). Normalization uncertainties of the data have
not been included. \label{fig:slac}}

\vspace{-0.5cm}

\end{figure}

We will also find it instructive to compare our results with the $F_{2}^{{\rm Fe}}/F_{2}^{{\rm D}}$
as extracted in neutral current charged-lepton scattering, $\ell^{\pm}$--$Fe$.
In Fig.~\ref{fig:slac} we compare the experimental results for the
structure function ratio $F_{2}^{{\rm Fe}}/F_{2}^{{\rm D}}$ for the
following experiments: \hbox{BCDMS-85} \citep{Bari:1985ga}, \hbox{BCDMS-87}
\citep{Benvenuti:1987az}, \hbox{SLAC-E049} \citep{Bodek:1983qn},
\hbox{SLAC-E139} \citep{Gomez:1993ri}, \hbox{SLAC-140} \citep{Dasu:1993vk}.
The curve (labeled SLAC/NMC parameterization) is a fit to this data.
While there is a spread in the individual data points, the parameterization
describes the bulk of the data at the level of a few percent or better.
It is important to note that this parameterization is independent
of atomic number $A$ and the energy scale $Q^{2}$ \citep{Arrington:2003nt};
this is in contrast to the results we will derive using the PDFs extracted
from the nuclear data.%
\footnote{In particular, we will find for large $x$ ($\gsim0.5$) and $Q$
comparable to the proton mass the target mass corrections for $F_{2}^{{\rm Fe}}/F_{2}^{{\rm D}}$
are essential for reproducing the features of the data; hence the
$Q$ dependence plays a fundamental role.%
} Additionally, we note that while this parameterization has been extracted
using ratios of $F_{2}$ structure functions, it is often applied
to other observables such as $F_{1,3,L}$ or $d\sigma$. We can use
this parameterization as a guide to judge the approximate correspondence
between this neutral current (NC) charged lepton DIS data and our
charged current (CC) neutrino DIS data.

\subsection{R Factors for $d^{2}\sigma/dx\, dQ^{2}$ \label{sec:xsec}}

\begin{figure}[t]
\includegraphics[width=0.95\columnwidth]{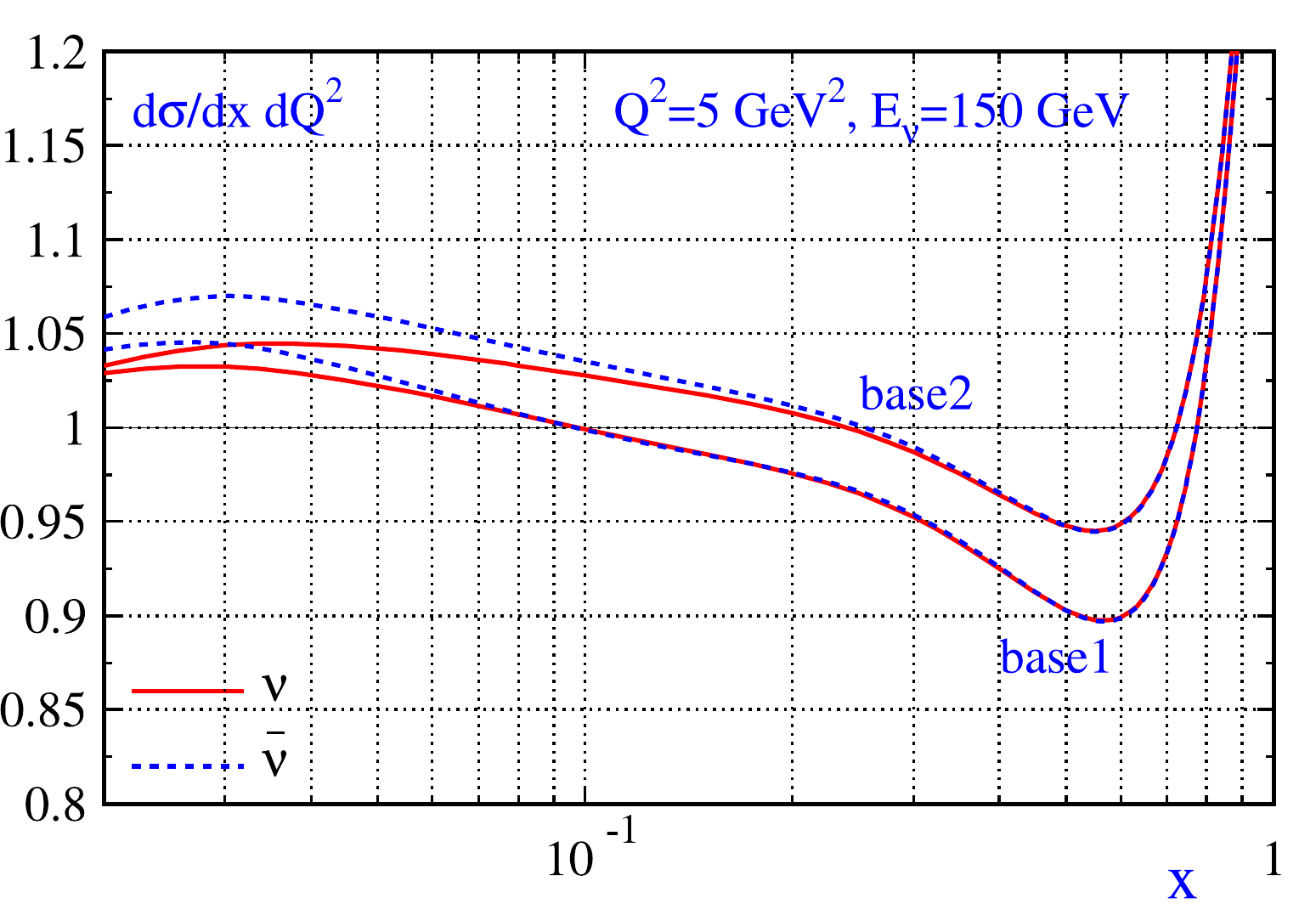}

\vspace{-1cm}
\caption{Nuclear correction factor $R$ for the differential cross section
$d^{2}\sigma/dx\, dQ^{2}$ in charged current neutrino-Fe scattering
at $Q^{2}=5~\gevsq$. Results are shown for the charged current neutrino
(solid lines) and anti-neutrino (dashed lines) scattering from iron.
The upper (lower) pair of curves shows the result of our analysis
with the Base-2 (Base-1) free-proton PDFs. \label{fig:dsig}}

\vspace{-0.5cm}

\end{figure}

We begin by computing the nuclear correction factor $R$ according
to Eq.\ (\ref{eq:R}) for the neutrino differential cross section
as this represents the bulk of the NuTeV data included in our fit.
More precisely, we show $R$-factors for the charged current cross
sections $d^{2}\sigma/dx\, dQ^{2}$ at fixed $Q^{2}$. Our results
are displayed in Fig.~\ref{fig:dsig} for $Q^{2}=5~\gevsq$ and a
neutrino energy $E_{\nu}=150\ \gev$ which implies, due to the relation
$Q^{2}=2ME_{\nu}xy$, a minimal $x$-value of $x_{{\rm min}}=0.018$.
The solid (dashed) lines correspond to neutrino (anti-neutrino) scattering
using the iron PDFs from the `A2' fit.

We have computed $R$ using both the Base-1 and Base-2 PDFs for the
denominator of Eq.~(\ref{eq:R}); recall that Base-1 includes a deuteron
correction while Base-2 uses the CCFR data and does not include a
deuteron correction. The difference between the Base-1 and Base-2
curves is approximately 2\% at small $x$ and grows to 5\% at larger
$x$, with Base-2 above the Base-1 results. The difference of these
curves, in part, reflects the uncertainty introduced by the proton
PDF.\citep{Schienbein:2007fs} As this behavior is typical, in the
following plots (Figs.~\ref{fig:fnu}) we will only show the Base-1
results. We also observe that the neutrino (anti-neutrino) results
coincide in the region of large $x$ where the valence PDFs are dominant,
but differ by a few percent at small $x$ due to the differing strange
and charm distributions.

\subsection{R Factors for $F_{2}^{\nu}(x,Q^{2})$ and $F_{2}^{\bar{\nu}}(x,Q^{2})$}

\begin{figure*}[t]
\begin{tabular}{|c|c|}
\hline 
\includegraphics[width=0.95\columnwidth]{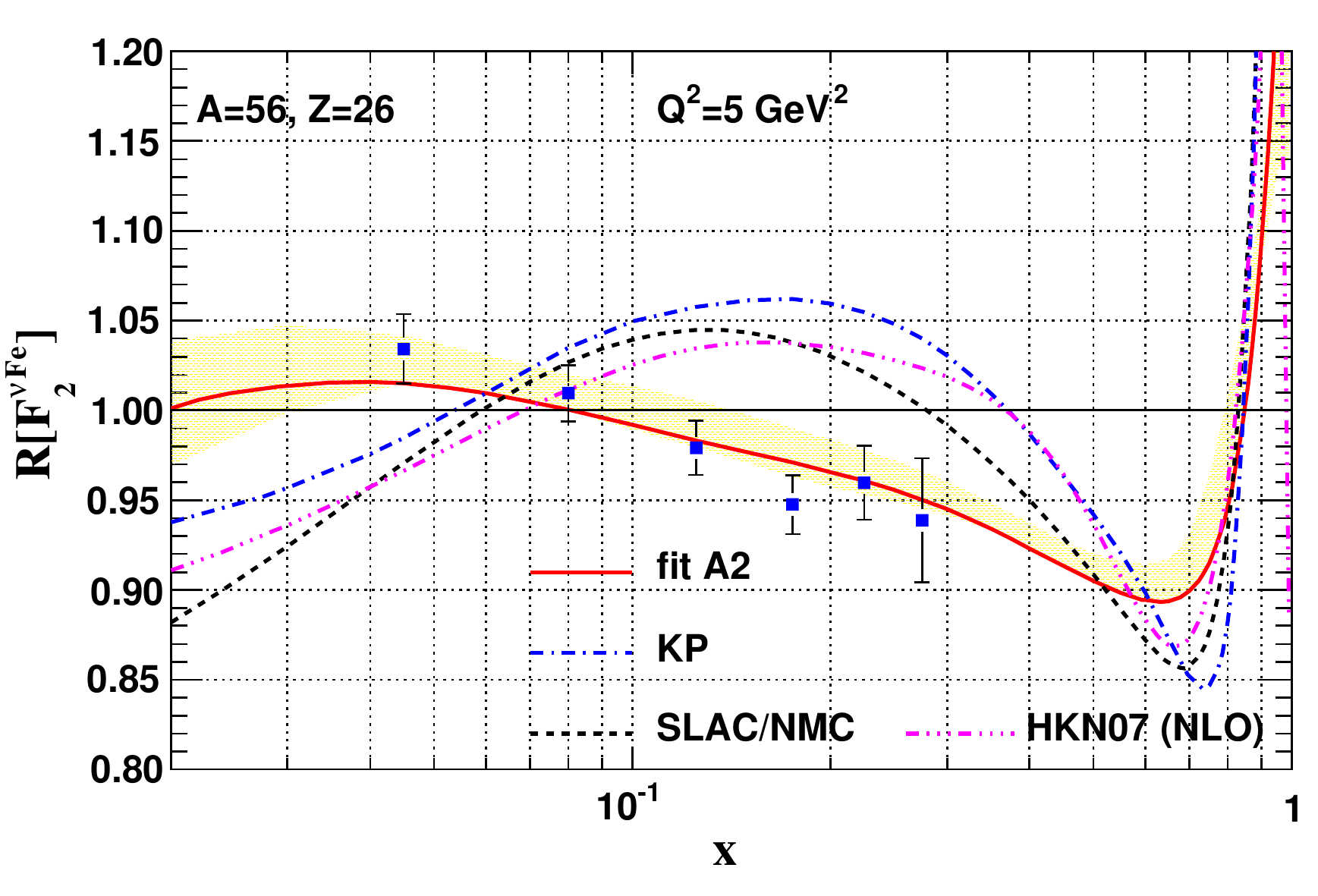} & \includegraphics[width=0.95\columnwidth]{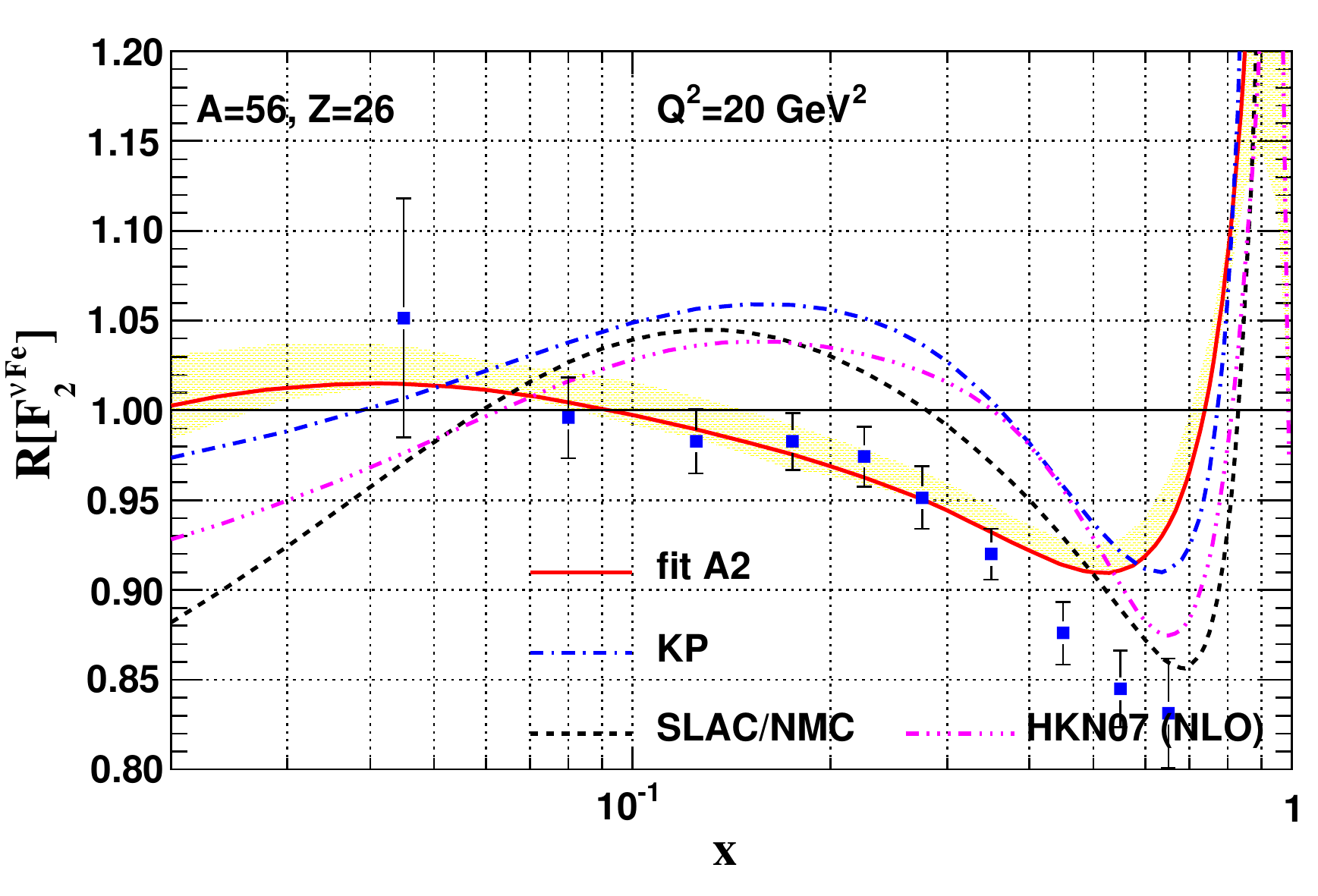}\tabularnewline
\hline
\hline 
\includegraphics[width=0.95\columnwidth]{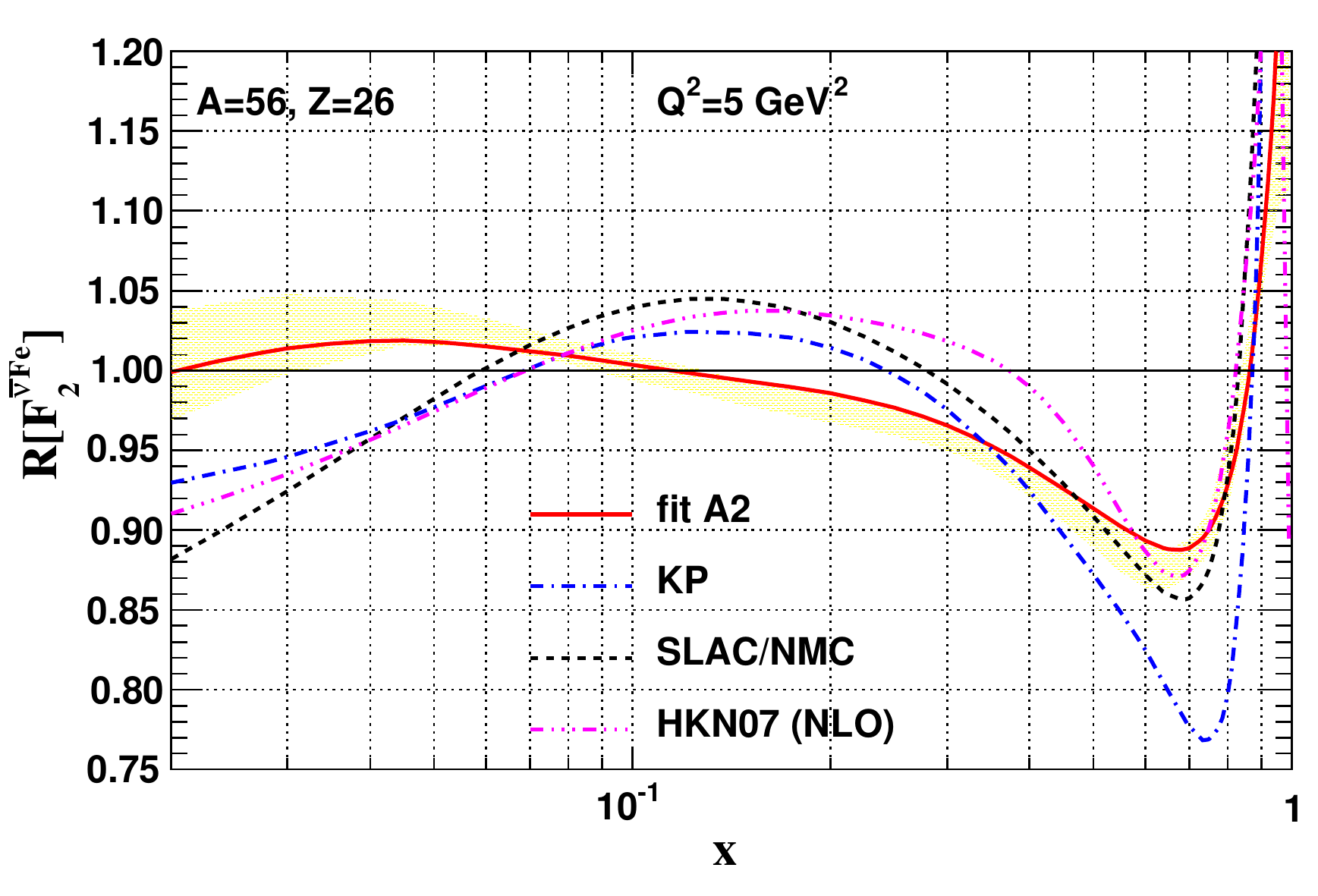} & \includegraphics[width=0.95\columnwidth]{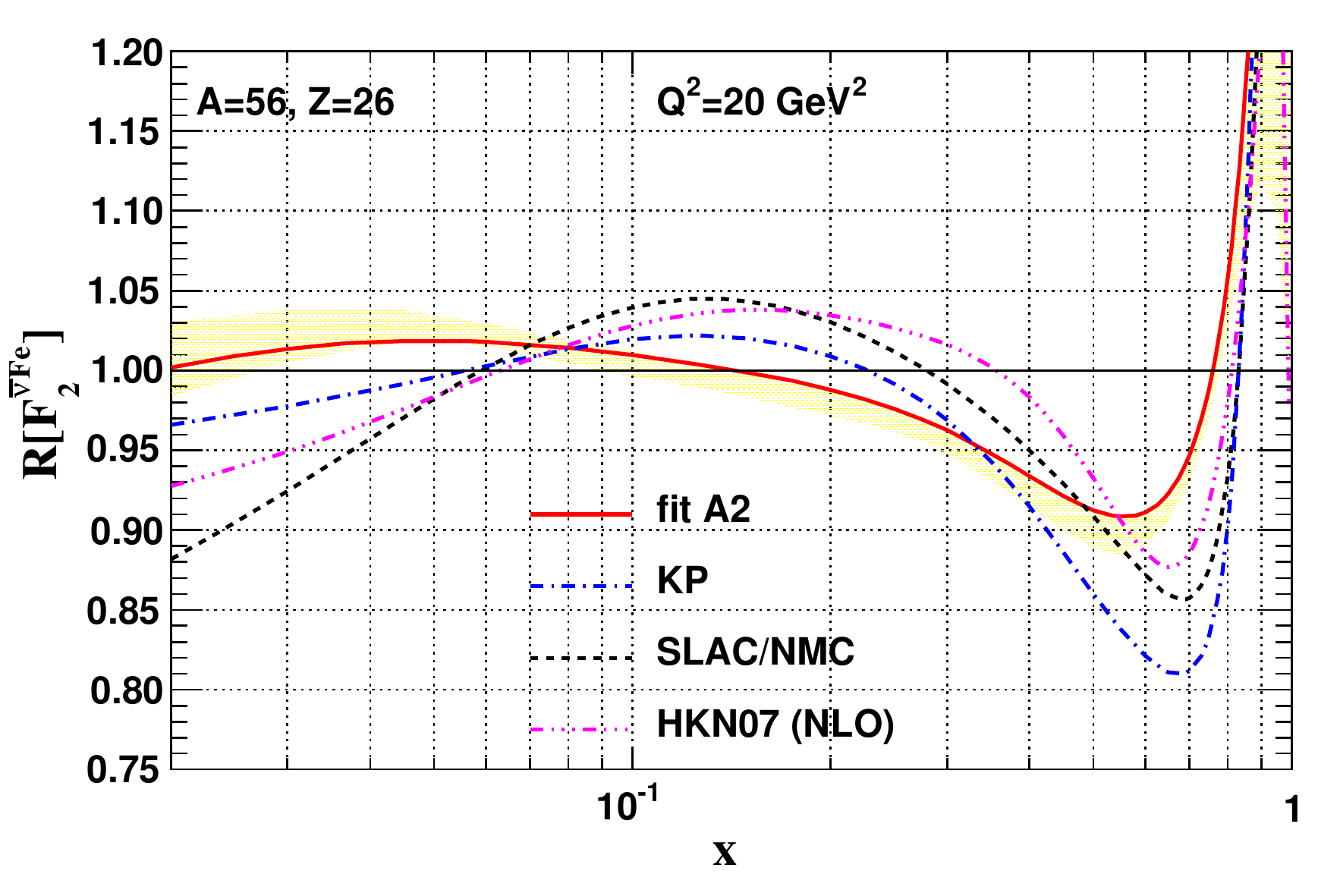}\tabularnewline
\hline
\end{tabular}

\vspace{-0.5cm}
\caption{Nuclear correction factor $R$ for the structure function $F_{2}$
in neutrino and anti-neutrino scattering from Fe for $Q^{2}=\{5,20\}$\,GeV$^{2}$.
The solid curve shows the result of our analysis of NuTeV data; the
uncertainty from the fit is represented by the shaded (yellow) band.
For comparison we show the correction factor from the Kulagin--Petti
model (dashed-dot line),\citep{Kulagin:2004ie} HKN07 (dashed-dotted
line),\citep{Hirai:2007sx} and the SLAC/NMC parametrization (dashed
line). \label{fig:fnu}}

\vspace{-0.25cm}

\end{figure*}

We now compute the nuclear correction factors for charged current
neutrino--iron scattering. The results for $\nu$--$Fe$ and those
of $\bar{\nu}$--$Fe$ are shown in Fig.~\ref{fig:fnu}. The numerator
in Eq.~(\eqref{eq:R}) has been computed using the nuclear PDF from
fit `A2', and for the denominator we have used the Base-1 PDFs. For
comparison we also show the correction factor from the Kulagin--Petti
model \citep{Kulagin:2004ie,Kulagin:2007ju} (dashed-dotted), and
the SLAC/NMC curve (dashed) which has been obtained from an $A$ and
$Q^{2}$-independent parameterization of calcium and iron charged--lepton
DIS data.

Due to the neutron excess in iron, both our curves and the KP curves
differ when comparing scattering for neutrinos and anti-neutrinos;
the SLAC/NMC parameterization is the same in both figures. For our
results (solid lines), the difference between the neutrino and anti-neutrino
results is relatively small, of order $3\%$ at $x=0.6$. Conversely,
for the KP model (dashed-dotted lines) the $\nu$--$\bar{\nu}$ difference
reaches $10\%$ at $x\sim0.7$, and remains sizable at lower values
of $x$.

To demonstrate the dependence of the $R$ factor on the kinematic
variables, in Fig.~\ref{fig:fnu} we have plotted the nuclear correction
factor for two separate values of $Q^{2}$. Again, our curves and
the KP model yield different results for different $Q^{2}$ values,
in contrast to the SLAC/NMC parameterization.

Comparing the nuclear correction factors for the $F_{2}$ structure
function with those obtained for the differential cross section (Fig.~\ref{fig:dsig}),
we see these are quite different, particularly at small $x$. Again,
this is because the cross section $d^{2}\sigma$ is comprised of a
different combination of PDFs than the $F_{2}$ structure function.
In general, our $R$-values for $F_{2}$ lie below those of the corresponding
$R$-values for the cross section $d\sigma$ at small $x$. Since
$d\sigma$ is a linear combination of $F_{2}$ and $F_{3}$, the $R$-values
for $F_{3}$ (not shown) therefore lie above those of $F_{2}$ and
$d\sigma$. Again, we emphasize that it is important to use an appropriate
nuclear correction factor which is matched to the particular observable.

As we observed in the previous section, our results have general features
in common with the KP model and the SLAC/NMC parameterization, but
the magnitude of the effects and the $x$-region where they apply
are quite different. Our results are noticeably flatter than the KP
and SLAC/NMC curves, especially at moderate-$x$ where the differences
are significant. The general trend we see when examining these nuclear
correction factors is that the anti-shadowing region is shifted to
smaller $x$ values and any turn-over at low $x$ is minimal given
the PDF uncertainties. In general, these plots suggest that the size
of the nuclear corrections extracted from the NuTeV data are smaller
than those obtained from charged lepton scattering (SLAC/NMC) or from
the set of data used in the KP model. We will investigate this difference
further in the following section.

\subsection{$F_{2}^{Fe}/F_{2}^{D}$ from Iron PDFs}

\begin{figure}[t]
\includegraphics[width=0.95\columnwidth]{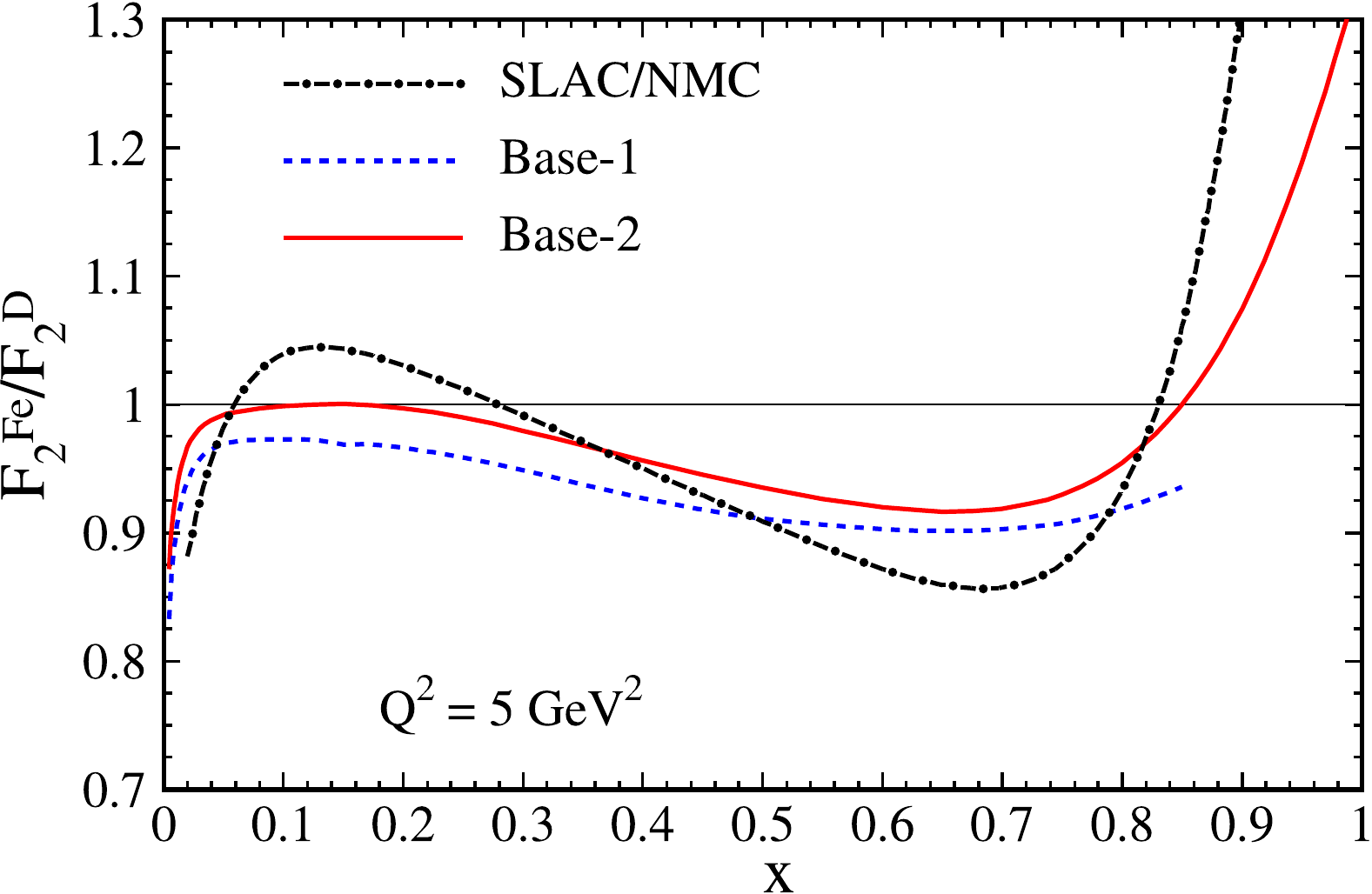}

\vspace{-0.5cm}
\caption{Predictions (solid and dashed line) for the structure function ratio
$F_{2}^{Fe}/F_{2}^{D}$ using the iron PDFs extracted from fits to
NuTeV neutrino and anti-neutrino data. The SLAC/NMC parameterization
is shown with the dot-dashed line. The structure function $F_{2}^{D}$
in the denominator has been computed using either the Base-2 (solid
line) or the Base-1 (dashed line) PDFs. \label{fig:future}}

\vspace{-0.5cm}

\end{figure}

Since the SLAC/NMC parameterization was fit to $F_{2}^{Fe}/F_{2}^{D}$
for charged-lepton DIS data, we can perform a more balanced comparison
by using our iron PDFs to compute this same quantity. The results
are shown in Fig.~\ref{fig:future} where we have used our iron PDFs
to compute $F_{2}^{Fe}$, and the Base-1 and Base-2 PDFs to compute
$F_{2}^{D}$.

As with the nuclear correction factor results of the previous section,
we find our results have some gross features in common while on a
more refined level the magnitude of the nuclear corrections extracted
from the CC iron data differs from the charged lepton data. In particular,
we note that the so-called {}``anti-shadowing'' enhancement at $x\sim[0.06-0.3]$
is \textit{not} reproduced by the charged current (anti-)neutrino
data. Examining our results among all the various $R[{\cal O}]$ calculations,
we generally find that any nuclear enhancement in the small $x$ region
is reduced and shifted to a lower $x$ range as compared with the
SLAC/NMC parameterization. Furthermore, in the limit of large $x$
($x\gtrsim0.6$) our results are slightly higher than the data, including
the very precise SLAC-E139 points; however,the large theoretical uncertainties
on $F_{2}^{D}$ in this $x$-region make it difficult to extract firm
conclusions.

This discussion raises the more general question as to whether the
charged current ($\nu$--$Fe$) and neutral current ($\ell^{\pm}$--$Fe$)
correction factors are entirely compatible \citep{Tzanov:2005kr,Tzanov:2005fu,Boros:1998qh,Boros:1999fy,Bodek:1999bb,Kretzer:2001mb}.
There is \textit{a priori} no requirement that these be equal; in
fact, given that the $\nu$--$Fe$ process involves the exchange of
a $W$ and the $\ell^{\pm}$--$Fe$ process involves the exchange
of a $\gamma$ we necessarily expect this will lead to differences
at some level.%
\footnote{In Ref.~\citep{Brodsky:2004qa}, Brodsky and collaborators posit
a non-universal nuclear anti-shadowing mechanism which yields different
effects for CC and NC scattering. %
}

\subsection{Future Studies \label{sec:summary}}

\begin{figure}[t]
\includegraphics[width=0.95\columnwidth]{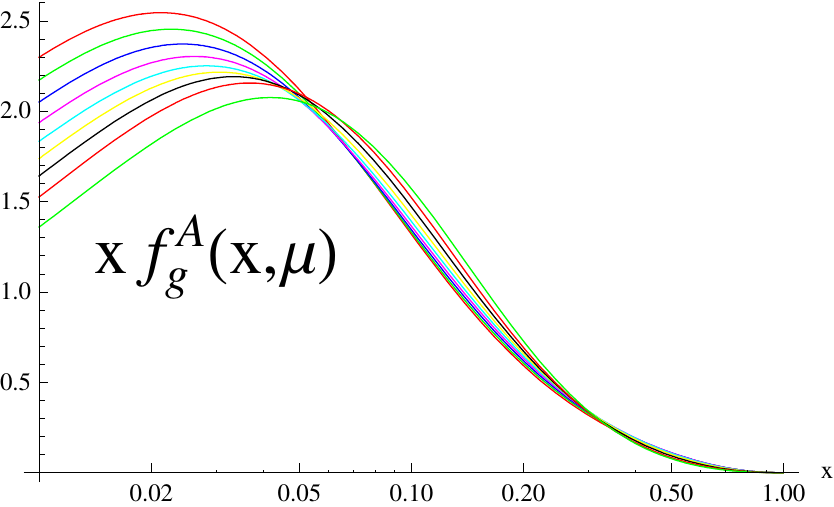}

\vspace{-0.5cm}
\caption{Illustration of the gluon PDF $x\, f_{g}^{A}(x,\mu)$ vs. $x$ as
a function of the nuclear $A$. \label{fig:gluon} }

\vspace{-0.5cm}

\end{figure}

It is important to resolve whether the differences we observe in Fig.~\ref{fig:future}
arise from the uncertainty of the nuclear corrections, or if they
are genuinely a consequence of NC/CC effects. A combined analysis
of CC neutrino and NC charged-lepton data sets will shed more light
on these issues. To best address these questions, we need to include
the nuclear dimension (parameterized by the nuclear A value) as a
dynamic component of the global fit; this will allow us to fit both
the CC $W^{\pm}$-exchange processes at large A, as well as the NC
$\gamma$-exchange processes at small A in a coherent framework. Figure~\ref{fig:gluon}
presents an illustrative example of how the PDFs can be extended to
incorporate the necessary A dependence to implement such a program.
This extended analysis with additional data sets is in progress, and
should help clarify these questions.

\section{CONCLUSIONS}

While the nuclear corrections extracted from charged current $\nu$--$Fe$
scattering have similar characteristics as the neutral current $l^{\pm}$--$Fe$
charged-lepton results, the detailed $x$ and $Q^{2}$ behavior is
quite different. This observation raises the deeper question as to
whether the charged current and neutral current nuclear correction
factors may be substantially different. This present study of the
iron PDFs provides a foundation for a general investigation (involving
a variable A parameter) that can definitively address this topic.
Resolving these questions is essential if we are to reliably use the
plethora of nuclear data to obtaining free-proton PDFs which form
the basis of the LHC analyses.

\section*{Acknowledgment}

We thank Tim Bolton, S. Brodsky, Javier Gomez, Shunzo Kumano, Eric
Laenen, Dave Mason, W. Melnitchouk, Donna Naples, Mary Hall Reno,
Voica~A.~Radescu, and Martin Tzanov for valuable discussions, and
BNL, CERN, and Fermilab for their hospitality. This work was supported
by U.S.\ DoE DE-FG02-04ER41299, DE-FG02-97IR41022, DE-AC05-06OR23177,
NSF grant 0400332, Lightner-Sams Foundation, \& Deutsche Forschungsgemeinschaft
(YU~118/1-1). 

\bibliographystyle{plunsrt}
\addcontentsline{toc}{section}{\refname}\bibliography{olnessNPDF}

\end{document}